\documentclass[aps,reprint,prd,amsmath,showpacs,showkeys,nobibnotes,nofootinbib]{revtex4-1}
\usepackage{amsfonts}
\usepackage{graphicx}
\DeclareMathAlphabet{\mathpzc}{OT1}{pzc}{m}{it}

\begin{document}
\title[Monopole and dipole layers]{Monopole and dipole layers in curved spacetimes: formalism and examples}

\date{\today}
\author{Norman {G{\"u}rlebeck}}
\email{norman.guerlebeck@gmail.com}
\author{Ji\v r\'i {Bi{\v c}{\'a}k}}
\email{jiri.bicak@mff.cuni.cz}

\affiliation{Institute of Theoretical Physics, Charles University, Prague, Czech Republic}

\author{Antonio C. \surname{Guti\'errez-Pi\~{n}eres}}
\email{acgutierrez@unitecnologica.edu.co}

\affiliation{Facultad de Ciencias B\'asicas, Universidad Tecnol\'ogica de Bol\'ivar, Cartagena de Indias, Colombia}

\begin{abstract}
The discontinuities of electromagnetic test fields generated by general layers of electric and magnetic monopoles and dipoles are investigated in general curved spacetimes. The equivalence of electric currents and magnetic dipoles is discussed. The results are used to describe exact ``Schwarzschild disk'' solutions endowed with such sources. 
The resulting distributions of charge and dipole densities on the disks are corroborated using the membrane paradigm.
\end{abstract}
\maketitle

\section{Introduction}

The investigation of electromagnetic fields coupled to strong gravitational fields have an interest from both theoretical perspectives and from a variety of applications in astrophysics. Examples on the theory side include studies of gravitational collapse of charged configurations (see, e.g., \cite{Kuchar_1968,Bicak_1972,*Bicak_1980a}), of the validity of the cosmic censorship conjecture \cite{Boulware_1973}, of the existence and properties of quasi black holes and wormholes (for recent accounts, see, e.g., \cite{Lemos_2010,Balakin_2010} and references therein), membranes producing repulsive gravity \cite{Belinski_2009}, and of many other issues. Very often analytical works employ, as tractable physical models, 2-dimensional thin shells sweeping out 3-dimensional timelike hypersurfaces. Recently, we used this idealization to construct ``spherical gravitating condensers'' -- two concentric charged shells made of perfect fluids (satisfying energy conditions) under the condition that the electric field is non-vanishing only between the shells (see \cite{Bicak_2010} and further references on charged shells therein).

The literature on electromagnetic fields in relativistic astrophysics\footnote{A very large number of papers is devoted to electromagnetic fields in cosmology -- both to more mathematical aspects like the Bianchi models with magnetic fields \cite{Wainwright_1997}, and to the more physical question of the origin of the fields \cite{Barrow_2007}.} is vast. Here we restrict ourselves to referring to several monographs dealing in detail with black-hole electrodynamics, e.g., \cite{Thorne_1986,Punsly_2008,Frolov_1998}, and we mention the relatively recent work \cite{Rezzolla_2001a,Rezzolla_2004} on electromagnetic fields around compact objects in which various papers are also summarized. In   \cite{Rezzolla_2001a,Rezzolla_2004} solutions to the Maxwell equations are presented both in the interior and outside a rotating neutron star and the matching conditions of the electromagnetic field at the stellar surface are analyzed in detail. The fields are not continuous across the stellar surface which gives rise to charges and currents. 

In the present paper we study electromagnetic sources distributed on shells in curved spacetimes in general, considering in particular possible discontinuities of the electromagnetic field across the shells. 
The sources discussed are layers with monopole or dipole currents. As far as we are aware electric or magnetic dipole layers and the matching conditions for their fields were not studied before in the context of general relativity. 

In general, in case of dipoles the currents and the electromagnetic field tensor will be distribution valued. This implies products of distributions in the stress-energy tensor. In order to avoid this, one can treat the electromagnetic field as a test field and solve the Maxwell equations in a given background metric. In many astrophysical situations this approach is well justified since typically the averaged energy density of the electromagnetic field is much smaller than that of the gravitational field. This approach is followed here and thus only the standard theory of generalized functions is used\footnote{If the full Einstein-Maxwell equations are to be solved, then the complicated formalism of generalized distributions, i.e., Colombeau algebras, might be used -- see, e.g., \cite{Grosser_2001,Steinbauer_2006}.}. 

In another work \cite{Gurlebeck_2011} we discuss spherical thin shells endowed with arbitrary, not necessarily spherical distributions of charge or dipole densities in a Schwarzschild spacetime. There it was possible to employ the results of \cite{Bicak_1977} to calculate the fields directly and read off their discontinuities across the shell. In the present paper we generalize the jump conditions to general backgrounds and general hypersurfaces. As a by-product of those jump conditions the equivalence of the external fields of magnetic dipoles and certain electric charge currents is proven in general. For elementary dipoles this was already known in special backgrounds like the Kerr spacetime \cite{Bicak_1976}. 

The jump conditions can be used to obtain massive disks endowed with charge and dipole densities using the Israel-Darboux formalism.
 In the examples studied here, we use the Schwarzschild disk spacetime as a background, cf. \cite{Bicak_1993}. Therefore, we generate massive thin disks (Schwarzschild disks) endowed with either electric/magnetic test charges or electric/magnetic test dipoles. The surface currents are depicted and explained using the membrane paradigm.

We use throughout the article the metric signature $(+1,-1,-1,-1)$ and units in which $c=G=1$.
 
\section{Monopole and dipole layers in general}

Although in the examples analyzed in section \ref{sec:examples} we use the Schwarzschild disk spacetimes as backgrounds, the results in the next section, i.e., the source terms and the jump conditions hold in a more general backgrounds. Of course, it has to admit a hypersurface, where the sources are situated, and the derivation of a dipole current requires, that a family of ``parallel'' hypersurfaces as defined below exists. In \cite{Kuchar_1968} the case of charged massive shells were already discussed in full Einstein-Maxwell theory. Nonetheless, we consider test charges in our work, mainly  to show in which cases the field generated by an arbitrary dipole distribution can be seen outside of the source as generated by moving charges.

\subsection{The 4-currents for charges or dipoles distributed on a shell}

Denoting by $F_{\alpha\beta}$ the Maxwell tensor and by ${}^*F_{\alpha\beta}$ its dual\footnote{\label{footnote:convention}Note that we use the signature $-2$ of the metric and the orientation of the volume form as in \cite{Jackson_1998}, with the important difference that the indices of our Maxwell tensor $F_{\alpha\beta}$ are interchanged.}, the Maxwell equations in a complex form read as follows:
\begin{align}\label{maxwell_equation}
  {\mathcal F}^{\alpha\beta}_{\phantom{\alpha\beta};\beta}=4\pi {\mathcal J}^\alpha,
\end{align}
where ${\mathcal F}_{\alpha\beta}=F_{\alpha\beta}+{\mathrm i}\, {}^*F_{\alpha\beta}$ is a self-dual 2-form. The 4-current ${\mathcal J}^\alpha=j^\alpha_{(e)}+{\mathrm i} j^\alpha_{(m)}$ consists of an electric part $j_{(e)}^\alpha$ and
a magnetic part $j_{(m)}^\alpha$. 
If $j^\alpha_{(m)}$ is vanishing the imaginary part of the Maxwell equations \eqref{maxwell_equation} allows us to introduce an electric 4-potential $A^{(e)}_\mu$ such that $F_{\mu\nu}= A^{(e)}_{\nu,\mu}-A^{(e)}_{\mu,\nu}$. In case there are no electric sources present, we can analogously introduce a magnetic 4-potential $ A^{(m)}_\mu$ such that $ {}^{*}F_{\mu\nu}= A^{(m)}_{\nu,\mu}-A^{(m)}_{\mu,\nu}$. In the vacuum region both 4-potentials can be defined and we denote the complex linear combination by $\mathcal A=A^{(e)}+{\mathrm i} A^{(m)}$.

Timelike hypersurfaces $\Sigma$ representing the history of charged 2-surfaces (shells) are discussed widely in the literature, see, e.g., \cite{Kuchar_1968}. We recall their main properties, in particular the form of the 4-current which will help us in formulating the expressions for the dipole current.
Suppose the hypersurface $\Sigma$ is described by $\Phi_\pm(x^\mu_\pm)=0$, where $x^\mu_\pm$ are coordinates in the two parts of the spacetime on the two sides of $\Sigma$ and the index $\pm$ denotes from which side a quantity is seen.
The unit normal of $\Sigma$ is given by $n_{\pm\mu}=\kappa N^{-1}_\pm \Phi_{\pm,\mu}|_{\Phi_\pm=0}$, where $N_{\pm}=(-\Phi_{\pm,\mu}\Phi_{\pm}^{,\mu}|_{\Phi_\pm=0})^{\tfrac 1 2}$ and $\kappa=\pm 1$ is chosen such that the normal points from $-$ to $+$.  
To shorten the notation we drop the index $\pm$ in the following wherever no confusion is to be expected. If the intrinsic coordinates of $\Sigma$ are called $\xi^a$, where $a$ runs from $0$ to $2$ and $\xi^0$ is a timelike coordinate, then the tangential vectors are $\mathrm{e}_a^\mu=\frac{\partial x^\mu}{\partial\xi^a}$. A tensor field $B_{\mu\ldots}$ can be projected onto these directions at $\Sigma$ and we denote this by
\begin{align}
 B_{a\ldots}=B_{\mu\ldots}\mathrm{e}^\mu_{a},\quad B_{\bot\ldots}=B_{\mu\ldots}n^\mu.
\end{align}

The 4-current of an electrically charged monopole layer is given by 
\begin{align}\label{eq:Monopolecurrent}
 j_{(e\mathpzc{Mo})}^{\mu}=s_{(e\mathpzc{Mo})}^a\mathrm{e}_a^\mu N \delta(\Phi), \quad s_{(e\mathpzc{Mo})}^a= \sigma_e u^a,
\end{align}
where $s_{(e\mathpzc{Mo})}^a$ is the surface current of the electrical charges, $\sigma_e$ is the rest electric surface charge density and $u^a$ the 4-velocity of the charged particles projected onto $\Sigma$. 

Let us consider, at least locally, a Gaussian normal coordinate system generated by the geodesics $\chi_p$ orthogonal to $\Sigma$ and going out from points $p\in\Sigma$. Then the metric is block diagonal 
\begin{align}\label{eq:gaussian}
 g_{\mu\nu}=-\left({\mathrm d}x^3\right)^2+\gamma^{(3)}_{ab}{\mathrm d}x^a{\mathrm d}x^b,
\end{align}
and $\Phi=x^3-x^3_0$.
The family of hypersurfaces $x^3=x^3_0+h$, i.e., $\Sigma(h)$, are still orthogonal to the family of geodesics ${\chi_p}$ and are at a proper distance $h$ measured along $\chi_p$ from $\Sigma(0)$. The coordinates $x^a$ can be used as intrinsic coordinates, thus $\mathrm{e}_a^\mu=\delta^\mu_a$ with the Kronecker delta $\delta^\mu_a$, and $\gamma_{ab}^{(3)}(x^a,x^3_0+h)$ denote the intrinsic metrics of the hypersurfaces $\Sigma(h)$ and $\gamma^{(3)}(x^a,x^3_0+h)$ their determinant. A slightly more general form of the metric arises when the coordinate $x^3$ along geodesics $\chi_p$ is not measuring anymore the proper distance, implying $g_{33}\neq 1$. These ``generalized'' Gaussian normal coordinates are used in the examples in section \ref{sec:examples}.
The 4-current of a charge distribution on the surface in the generalized Gaussian normal coordinates is given by
\begin{align}
  j_{(e\mathpzc{Mo})}^{\mu}=s_{(e\mathpzc{Mo})}^a\mathrm{e}_a^\mu \left(-g_{33}(x^a,x^3_0)\right)^{-\frac 1 2}\delta(x^3-x^3_0).
\end{align}
To avoid confusion, we recall the definition of the 1-dimensional $\delta$-distribution: For any sufficiently smooth test function $f$ the following integral over a spacetime region $\Omega$ in the generalized Gaussian normal coordinates reduces to the integral over $\Sigma(0)$ as follows:
\begin{align}
\begin{split}
&\int\limits_\Omega f(x^a,x^3_0)\left(-g_{33}(x^a,x^3_0)\right)^{-\frac 1 2} \delta(x^3-x^3_0)\mathrm{d}\Omega=\\
&\int\limits_{\Sigma\cap \Omega}f(x^a,x^3_0)\mathrm{d}\Sigma,
\end{split}
\end{align}
where $\mathrm{d} \Omega$ is a spacetime volume element, $\mathrm{d}\Sigma$ is a volume element of the hypersurface $\Sigma(0)$.

\begin{figure}
\includegraphics[scale=0.55]{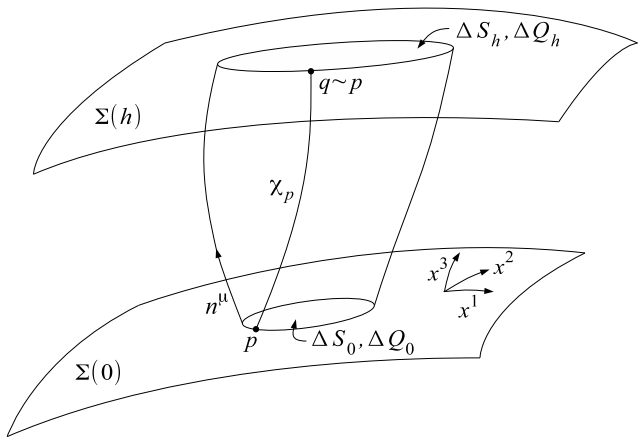}
\caption{\label{Fig:Volumeelement}Equivalent points and associated 3-volumes. The timelike coordinate $x^0$ is suppressed.}
\end{figure}

\begin{figure}
\includegraphics[scale=0.55]{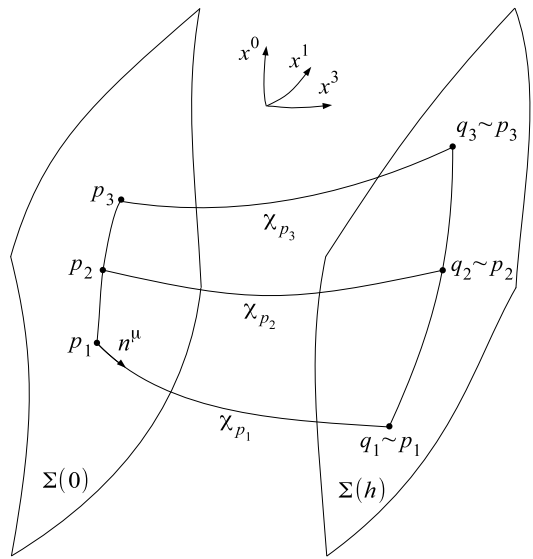}
\caption{\label{Fig:Motion}The motion of two associated infinitesimal charges. The spacelike coordinate $x^2$ is suppressed.}
\end{figure}

We construct dipole layers from two oppositely charged monopole layers which are separated by a proper distance $h$. For simplicity we derive the 4-current in the Gaussian normal coordinates \eqref{eq:gaussian} and make a coordinate transformation to the generalized Gaussian normal coordinates subsequently. Dipole layers arise in the limit of vanishing $h$ with a simultaneous limit to infinite (and opposite) rest surface charge densities of the two shells. This means we consider two hypersurfaces $\Sigma(0)$ and $\Sigma(h)$ endowed with surface rest charge densities of opposite sign\footnote{The second argument of the function $\sigma_h(\xi^a;x^3)$ denotes the layer $\Sigma_{x^3-x_0^3}$ on which the current is given and the index $h$ labels different currents during the limiting procedure -- the increase/decrease of the charge densities while bringing both shells together.} $\sigma_h(x^a;x^3_0)$ and $\sigma_h(x^a;x^3_0+h)$ and with velocity fields  $u^\mu(x^a;x^3_0)$ and $u^\mu(x^a;x^3_0+h)$, so giving rise to two 4-currents. Note that the change of the charge densities in the limit is such that it does not effect the velocity fields. Dipole layers result only in the limiting process $h\to 0$ if certain properties hold true in the limit which, for simplicity, we assume to hold throughout the entire limiting procedure. The family of geodesics $\chi_p$ gives locally rise to an equivalence relation of points similarly to \cite{Geroch_1971}, i.e., $p\sim q$ if there exist a point $p_0$ such that $p,~q\in\chi_{p_0}$, cf. Fig. \ref{Fig:Volumeelement}.
Since the intrinsic coordinates are carried along the geodesics, equivalent points are characterized by the same intrinsic coordinates. 
Let us assume that two charge elements initially placed at two equivalent points $(\xi^a,x_0^3)$ and $(\xi^a,x_0^3+h)$ stay in course of there motion in equivalent points at every moment of time, e.g., the intrinsic time coordinate $x^0$, cf. Fig. \ref{Fig:Motion}. Then the coordinate velocities of the charge elements are the same, so we have
\begin{align}\label{eq:conditionvelocity}
 \frac{u^\mu(\xi^a;x_0^3)}{u^0(\xi^a;x^3_0)}=\frac{u^\mu(\xi^a;x_0^3+h)}{u^0(\xi^a;x^3_0+h)}.
\end{align}
Analogously to the the equivalence of points, we also can, with each area element $\Delta S_0$ in $\Sigma(0)$, associate an area element $\Delta S_h$ in $\Sigma(h)$ which is cut out by the geodesics emanating from the boundary of $\Delta S_0$, cf. Fig. \ref{Fig:Volumeelement}. Since the total charge of the dipole shell has to vanish, we suppose that the charge $\Delta Q_0$ enclosed in any area $\Delta S_0$ (as seen by observers at rest with respect to the intrinsic coordinates) is the opposite of that enclosed in $\Delta S_h$. This condition yields
\begin{align}\label{eq:chargedensityratio}
\frac{\sigma_h(\xi^a;x^3_0) u^0(\xi^a;x^3_0)}{\sqrt{\gamma^{(3)}(x^a,x^3_0+h)}} =-\frac{\sigma_h(\xi^a;x^3_0+h)u^0(\xi^a;x^3_0+h)}{\sqrt{\gamma^{(3)}(x^a,x^3_0)}}.
\end{align}
As in the classical case, the charge density $\sigma_h(x^a;x^3_0)\to \pm\infty$ as $h\to 0$. The electrical rest dipole moment surface density is then naturally defined as $d_e(x^a)=-\lim\limits_{h\to 0}\sigma_h(x^a;x^3_0)\, h$.

Therefore, the limiting procedure based on \eqref{eq:Monopolecurrent}, \eqref{eq:conditionvelocity} and \eqref{eq:chargedensityratio} yields the resulting dipole 4-current in the form
\begin{align}
  j_{(e\mathpzc{Di})}^{\mu}=-d_e(x^a) u^\mu(x^a) \frac{\sqrt{\gamma^{(3)}(x^a,x^3_0)}}{\sqrt{\gamma^{(3)}(x^a,x^3)}}\,\delta'(x^3-x^3_0).
\end{align}
Of course, the total charge contained in any proper volume enclosing a part of the electric dipole layer is vanishing. 
Rewriting this in the generalized Gaussian normal coordinate system we find the 4-current to read
\begin{align}\label{eq:dipoledensity}
\begin{split}
 j_{(e\mathpzc{Di})}^{\mu}&=s_{(e\mathpzc{Di})}^a\mathrm{e}_a^\mu \left. \frac{\sqrt{\gamma^{(3)}}}{\sqrt{-g_{33}}}\right|_{x^3=x^3_0} \frac{1}{\sqrt{-g}} \delta'(x^3-x^3_0),\\
s_{(e\mathpzc{Di})}^a&= -d_e u^a,
\end{split}
\end{align}
where $s_{(e\mathpzc{Di})}^a$ is the surface current of the electrical dipoles and $u^a$ is the 4-velocity of the dipoles projected onto $\Sigma$. 
Let us repeat the definition of the normal derivative of a $\delta$-distribution in a curved background. For an arbitrary, sufficiently smooth test function $f$ the following holds: 
\begin{align}
\begin{split}
&\int\limits_\Omega f(x^a,x^3)\left. \frac{\sqrt{\gamma^{(3)}}}{\sqrt{-g_{33}}}\right|_{x^3=x^3_0} \frac{1}{\sqrt{-g}}\delta'(x^3-x^3_0)\mathrm{d}\Omega=\\
&-\int\limits_{\Sigma\cap \Omega}(n^\mu f_{,\mu})(x^a,x^3_0)\mathrm{d}\Sigma.
\end{split}
\end{align}

Note that even though a derivative of the delta function appears, no metric functions have to be differentiated because of the integral definition of distributions where $\sqrt{-g}$ appears and cancels with the only metric term in the 4-current depending on $x^3$. Thus also metrics which are not $C^1$ as they arise in the Israel formalism are allowed. It is also clear by construction and a short calculation, that the continuity equation for $j^\mu$ implies that the surface currents $s_{(e\mathpzc{Mo})}^a$ and $s_{(e\mathpzc{Di})}^a$ satisfy the continuity equation on $\Sigma$.
The currents for shells endowed with a magnetic charge or a magnetic dipole density are analogously defined, i.e., we just have to replace the index $e$ by the index $m$.

\subsection{Discontinuities in the potential and the fields}\label{subsec:DiscontinuitiesGeneral}

As is well known from flat space, the jumps of various components of the fields or potentials across a surface are related to electromagnetic sources distributed on that surface. However, even in special relativity magnetic charges are usually not discussed.
The jumps resulting from a dipole layer were, to the best of our knowledge, not discussed in curved spacetimes. We denote the jumps of a function $f$ by $[f]=f_{+}-f_{-}$. We study the four cases of electric/magnetic charged shells and electric/magnetic dipole shells separately. All of them can be obtained using the equivalence principle and Maxwell theory. 

In the case of  an electrically charged surface Kucha\v r showed in \cite{Kuchar_1968} (see also \cite{Israel_1966,Cruz_1967,Barrabes_1991}) that\footnote{The differences in the sign have their origin in a different signature of the metric.}
\begin{align}\label{eq:discontinuity_electrical_monopole}
 [F_{(e\mathpzc{Mo}) a \bot}]=-4\pi s_{(e\mathpzc{Mo}) a}, \quad [F_{(e\mathpzc{Mo})a b}]=0.
\end{align}
Note that these equations are covariant with respect to a change of intrinsic coordinates $\xi^a$ and scalars with respect to the coordinates $x^\mu$.
For the electric 4-potential in an appropriate Lorenz gauge it follows
\begin{align}\label{eq:discontinuity_potential_eM}
[A^{(e)}_{(e\mathpzc{Mo}) a}]= [A^{(e)}_{(e\mathpzc{Mo}) \bot}]=0.
\end{align}
The magnetic 4-potential $A^{(m)\mu}$ will in general not be continuous across $\Sigma$ owing to the fact that it can only be introduced in the absence of electrical currents and, therefore, different potentials will occur in the lower and the upper half of the spacetime. Furthermore, introducing the potential $A^{(e)\mu}$ on both sides of $\Sigma$ in different gauges will not change the external field, however, jumps in the potential are, as seen below, related to dipole densities and therefore describe a different physical system; in particular, the field {\em{in}} $\Sigma$ is changed. 

In case of a shell endowed with magnetic charges the same equations as \eqref{eq:discontinuity_electrical_monopole} and \eqref{eq:discontinuity_potential_eM} hold for the dual of the Maxwell tensor and for the magnetic 4-potential in a Lorenz gauge:
\begin{align}\label{eq:discontinuity_magnetic_monopole}
[{}^*F_{(m\mathpzc{Mo}) a \bot}]&=-4\pi s_{(m\mathpzc{Mo}) a}, \quad [{}^*F_{(m\mathpzc{Mo})a b}]=0,\notag\\
[A^{(m)}_{(m\mathpzc{Mo}) a}]&= [A^{(m)}_{(m\mathpzc{Mo}) \bot}]=0.
\end{align}
For the Maxwell tensor it follows that the tangential components jump and the normal components are continuous:
\begin{align}
[F_{(m\mathpzc{Mo}) a \bot}]&=0,\quad [F_{(m\mathpzc{Mo}) a b}]= 4\pi \epsilon^{(3)}_{abc}{s_{(m\mathpzc{Mo})}^c},
\end{align}
where $\epsilon^{(3)}_{abc}=\epsilon_{abc\bot}$ is the volume form of $\Sigma$ related to the induced metric $\gamma_{ab}$ of $\Sigma$ whereas $\epsilon_{\alpha\beta\gamma\delta}$ is the volume form of the spacetime. Tangential indices are raised and lowered with the induced metric and its inverse.

Analogously, from the equivalence principle the discontinuities of the Maxwell tensor for electric and magnetic dipole densities follow:
\begin{align}\label{eq:discontinuity_dipole}
 [F_{(e\mathpzc{Di}) a\bot}]&=0,& [F_{(e\mathpzc{Di})a b}]&= -8\pi {s_{(e\mathpzc{Di})[a,b]}},\notag\\
[{}^*F_{(m\mathpzc{Di}) a\bot}]&=0,& [{}^*F_{(m\mathpzc{Di})a b}]&= -8\pi {s_{(m\mathpzc{Di})[a,b]}}.
\end{align}
Here the antisymmetrization in the derivatives of $s_a$ is defined as $B_{[ab]}=\tfrac 12(B_{ab}-B_{ba})$. Note that a layer with a curl-free $s_{(m\mathpzc{Di})a}$ will not produce a jump in the external field and thus the source can only be detected by observing the trajectories of particles crossing that layer, i.e., by measuring the internal field in $\Sigma$.

The 4-potentials satisfy in these cases the following jump conditions:
\begin{align}\label{eq:jump_in_potential_DP}
[A^{(e)}_{(e\mathpzc{Di}) \bot}]&=0,&[A^{(e)}_{(e\mathpzc{Di}) a}]&=4\pi s_{(e\mathpzc{Di}) a},\\
[A^{(m)}_{(m\mathpzc{Di}) \bot}]&=0,& [A^{(m)}_{(m\mathpzc{Di}) a}]&=4\pi s_{(m\mathpzc{Di}) a}.
\end{align}
Additionally, the normal components of the Maxwell tensor have a $\delta$-like contribution $V_{a}=[A_a]N\delta(\Phi)$, the field ``between the two layers''. In order to see this contribution, we insert the aforementioned jumps into the Maxwell equations and calculate the source. Using again Gaussian normal coordinates we start with an electric 4-potential which is discontinuous across $\Sigma$ and calculate the sources. Hence, we write
\begin{align}
 A^{(e)}_\mu=A^{(e)+}_\mu \theta(x^3-x^3_0)+A^{(e)-}_\mu \theta(-x^3+x^3_0),
\end{align}
with $[A^{(e)}_z]=0$, which implies the Maxwell tensor to be
\begin{align}\label{eq:faradaytensor}
\begin{split}
 F_{a \mu}= &F^{+}_{a\mu} \theta(x^3-x^3_0)+F^{-}_{a\mu} \theta(-x^3+x^3_0)\\
-&\delta^z_\mu [A^{(e)}_a]\delta(x^3-x^3_0).
\end{split}
\end{align}
Inserting this into the Maxwell equations and using the jump conditions above yields
\begin{align}
\begin{split}
 F^{\mu \nu}_{\phantom{\mu\nu};\nu}=&4\pi s^a_{(e\mathpzc{Mo})} \mathrm{e}^\mu_a\delta(x^3-x^3_0)+\\
&4\pi s^a_{(e\mathpzc{Di})} \frac{\sqrt{\gamma^{(3)}(\xi^a,x^3_0)}}{\sqrt{\gamma^{(3)}(\xi^a,x^3)}}\mathrm{e}^\mu_a\delta'(x^3-x^3_0)+\\
&4\pi j^\mu_{+} \theta(x^3-x^3_0)+4\pi j^\mu_- \theta(x^3_0-x^3),
\end{split}
\end{align}
where the first two terms are the source terms for a charged layer and for a dipole layer. The last two terms are sources outside of $\Sigma$, for instance a volume charge density. In the remainder we will assume that outside of the shell there are no magnetic or electric sources.

\subsection{The equivalence of electric charges and magnetic dipoles}

In flat spacetimes and also in certain cases of electromagnetism in curved backgrounds, e.g., in the Schwarz\-schild and the Kerr spacetimes \cite{Bicak_1976,Bicak_1977}, 
the equivalence of the external field of a magnetic point dipole and of an infinitesimal 
electric charge current loop is known and often used. Naturally, it can also be easily shown that the external field of an electric point
dipole is indistinguishable from that of an infinitesimal magnetic charge current loop. A similar result can be shown to hold in the case of layers of dipoles. 
In our Gaussian normal coordinates the dual of the Maxwell tensor for a shell endowed with magnetic dipoles 
reads as follows, cf. \eqref{eq:jump_in_potential_DP} and \eqref{eq:faradaytensor}:
\begin{align}\label{eq:faradaymagneticdipoles}
  {}^{*}F^{(m\mathpzc{Di})}_{a \mu}=&   {}^{*}F^{(m\mathpzc{Di})+}_{a\mu} \theta(x^3-x^3_0)+  {}^{*}F^{(m\mathpzc{Di})-}_{a\mu} \theta(x^3_0-x^3)\notag\\
	&-4\pi \delta^z_\mu s^{(m\mathpzc{Di})}_a\delta(x^3_0).
\end{align}
Of course, the internal field must be changed to transform locally from sources in the form of magnetic dipoles to electric currents. However, if we remove the last term in \eqref{eq:faradaymagneticdipoles} from the field the external field remains unchanged. An observer outside can detect the difference only by examining trajectories of charged test particles crossing the shell. Furthermore, the jumps of the Maxwell tensor remain the same:
\begin{align}
 [F_{(m\mathpzc{Di})a\bot}]=4\pi \epsilon^{(3)\phantom a bc}_{\phantom{(3)}a}s_{(m\mathpzc{Di}) b,c},\quad [F_{(m\mathpzc{Di})ab}]=0.
\end{align}
Using equation \eqref{eq:discontinuity_electrical_monopole}, these jumps are produced by an electric current $s^a_{(e\mathpzc{Mo})}$ if 
\begin{align}\label{eq:equivalence}
s^a_{(e\mathpzc{Mo})}=-\epsilon^{(3) a b c} s_{(m\mathpzc{Di})b,c}.
\end{align}

The electric charge current defined in such a way can also be seen as a source. The continuity equation for $s_{(e\mathpzc{Mo})}^a$ is satisfied trivially.
However, since the charge density $s_{(e\mathpzc{Mo})}^0$ does not need to vanish, electrical charges are introduced in general. The total charge is in principle detectable at infinity in the asymptotics of the field assuming it falls off sufficiently fast. 
Nonetheless, the total charge for a field generated by magnetic dipoles is vanishing. How is this to be resolved?
The total electric charge $Q$ of $\Sigma$ as seen for observers at rest with respect to the intrinsic coordinates is given by
\begin{align}\label{eq:totalcharge}
 Q=\int\limits_{\Sigma\cap\{x^0=x^0_0\}}\left.s^0_{(e\mathpzc{Mo})}\sqrt{\gamma^{(3)}(\xi^a,x_0^3)}\right|_{\xi^0=x^0_0}\mathrm{d}\xi^1\mathrm{d}\xi^2.
\end{align}
Together with equation \eqref{eq:equivalence} and Stokes' theorem we obtain
\begin{align}\label{eq:line_integral}
Q=\int\limits_{\partial(\Sigma\cap\{x^0=x^0_0\})}(s_{(m\mathpzc{Di})1}\mathrm{d}\xi^1+s_{(m\mathpzc{Di})2}\mathrm{d}\xi^2).
\end{align}
The asymptotic behavior of the field implies a vanishing current at infinity\footnote{For closed shells this integral vanishes trivially.}. Thus, no {\em{total}} electric charge $Q$ will be present though ``local'' volumes can contain a net charge. This is also in correspondence with the known results for point dipoles. In a rest frame of a point dipole the external field can be seen as caused by an infinitesimal charge current loop with a vanishing time component. This is usually interpreted as two currents of positive and negative charges such that the charge densities in the rest frame of the dipole cancel each other and -- for example the positive charges are at rest (ions of the conductor) and the negative charges (electrons) contribute to the current. However, in a general frame as used here the charge densities do not necessarily cancel anymore. To generalize this to layers these point dipoles have to be superposed and so the current loops. The net current can have a charge density because one is not in a comoving frame of the dipoles.

If the fields do not fall off sufficiently fast, then the total charge of the shell need not vanish or be definable. In such a case charges can also be ``placed at infinity" which is reflected by a corresponding boundary condition. An example is given in section \ref{subsec:Asymptoticallyhomogeneous}.

The argument given above can be reversed and used to show that the external field of every electric charge surface current can also be produced by a charge density at rest in a given frame of reference and a magnetic dipole surface current.  The integrability condition of equation \eqref{eq:equivalence} for $s^a_{(m\mathpzc{Di})}$ is then equivalent to the continuity equation of the electric charge surface current. It is obvious that an analogous equivalence between electric dipoles and magnetic charges can be established.
Except for this kind of non-uniqueness, the field and its sources are completely determined by the jump conditions \eqref{eq:discontinuity_electrical_monopole}-\eqref{eq:jump_in_potential_DP}. 

\section{Schwarzschild disks with electric/magnetic charge and dipole density}\label{sec:examples}

The Schwarzschild metric in Schwarzschild coordinates $(x^\mu)=(t,r,\theta,\varphi)$ reads 
\begin{align}
\begin{split}
\mathrm{d}s^2 =& \left(1 - \frac{2M}{r}\right){\mathrm d}t^2 - \left(1 - \frac{2M}{r}\right)^{-1}{\mathrm d}r^2\\
& - r^2\left({\mathrm d}\theta^2 + \sin^2\theta {\mathrm d}\varphi^2 \right).
\end{split}
\end{align}
In \cite{Bicak_1993} massive disks of counterrotating matter, the ``Schwarzschild disks'', were constructed from this spacetime
using the Israel-Darboux formalism and Weyl coordinates $(x^\mu)=(t,\rho,z,\varphi)$
\begin{align}\label{coord_trafo}
\rho=\sqrt{r^2-2Mr}\sin\theta,\quad z=(r-M)\cos \theta.
\end{align}
This was done by identifying the surfaces $z= z_0$ and $z= -z_0$. From the jumps of the extrinsic curvature of the resulting surface an energy-momentum density of the disk was obtained. The disks are infinite but their mass is finite and the mass density decreases rapidly at large radii.
We show here how to endow such disks with an electric/magnetic charge densities or electric/magnetic dipole densities in a test field approach. We demonstrate this with two examples using the asymptotic homogeneous field and the field generated by a point charge. The same can be done to model more general distributions using the general solutions of the Maxwell equations for test fields on a Schwarzschild background given in \cite{Bicak_1977}. 

In $\Sigma$ defined by $z=z_{\pm 0}$, we introduce intrinsic coordinates $(\xi^0,\xi^1,\xi^2)=(T,R,\varPhi)$ which coincide with the Schwarzschild coordinates $(t,r,\varphi)$ in the disk but are capitalized to prevent confusion. The components of the normal vector in Schwarzschild coordinates are given by
\begin{align}
\begin{split}
(n_\mu)&=N (0,\cos\theta_{\pm},-(R-M)\sin\theta_+,0)\\
N=&-\left(1-\frac{2M}{R} +\frac{M^2}{R^2}\sin^2\theta_+\right)^{-\tfrac 1 2},
\end{split}
\end{align}
where again ``$+$'' denotes the quantities as seen from $z>z_0$ and ``$-$'' as seen from $z<-z_0$. Note that $\cos\theta_\pm=\pm \tfrac{z_0}{R-M}$.

\subsection{Asymptotically homogeneous electric and magnetic field}\label{subsec:Asymptoticallyhomogeneous}

The first test field to be discussed is the asymptotically homogeneous electric and magnetic field, for which the complex 4-potential and Maxwell tensor in Schwarzschild coordinates read as follows (see, e.g., \cite{Bicak_1977})
\begin{align}\label{asymptotically_homogeneous}
\begin{split}
 {\mathcal A}_t&=-{\mathcal F}_0 (r-2M)\cos\theta+{\mathcal A}_{t0},\\
 {\mathcal A}_\varphi&=-\frac{\mathrm i}{2}{\mathcal F_0}\sin^2\theta r^2+\mathcal A_{\varphi 0},\\
 {\mathcal A}_r&={\mathcal A}_\theta=0,\\
 {\mathcal F}_{tr}&={\mathcal F}_0\cos \theta,\\
 {\mathcal F}_{t\theta}&=-{\mathcal F}_0(r-2M)\sin \theta,\\
 {\mathcal F}_{\theta\varphi}&=-{\mathrm i}{\mathcal F}_0 r^2 \cos \theta \sin \theta,\\
 {\mathcal F}_{r\varphi}&=-{\mathrm i}{\mathcal F}_0 r \sin^2 \theta,\\
 {\mathcal F}_{t\varphi}&={\mathcal F}_{r\theta}=0,\\
 {\mathcal F}_0&=E_0+{\mathrm i} H_0.\\
\end{split}
\end{align}
The 4-potential is in fact not given in \cite{Bicak_1977} but can be calculated easily.
Assume the field in the upper/lower half is parametrized by ${\mathcal F}_{0\pm},~{\mathcal A}_{t0\pm}$ and ${\mathcal A}_{\varphi 0\pm}$.
The jumps of the potential across $\Sigma$ are given by
\begin{align}\label{eq:jumppotentialhomogeneous}
\begin{split}
 [{\mathcal A}_T]&=-(R-2M)\cos\theta_+ ({\mathcal F}_{0+}+{\mathcal F}_{0-})+\mathcal A_{T 0 +}-\mathcal A_{T 0 -},\\
[{\mathcal A}_R]&=[{\mathcal A}_\bot]=0,\\ 
[{\mathcal A}_\varPhi]&=-\frac{\mathrm i}{2}\sin^2\theta_+R^2 ({\mathcal F}_{0+}-{\mathcal F}_{0-})+\mathcal A_{\varPhi 0 +}-\mathcal A_{\varPhi 0 -}.
\end{split}
\end{align}
As it should be according to the equations \eqref{eq:discontinuity_potential_eM}, \eqref{eq:discontinuity_magnetic_monopole} and \eqref{eq:jump_in_potential_DP}, the orthogonal component of the potential is continuous. Furthermore, the radial component is continuous as well, i.e., the dipole currents (electric or magnetic) in the radial direction are vanishing. The dipole density approaches a constant value, so does the current in the $\varPhi$ direction, as one can expect from the analogous result obtained in Maxwell theory in flat space or after setting the mass $M$ to zero in the equations above.
The jumps in the fields read
\begin{align}
\begin{split}
 [{\mathcal F}_{T \bot}]&=N\left(1-\frac{2M}{R}\right)\left(1-\frac{M}{R} \sin^2\theta_+\right)  \left({\mathcal F}_{0-}-{\mathcal F}_{0+}\right),\\
 [{\mathcal F}_{\varPhi \bot}]&={\mathrm i} N M\cos\theta_+\sin^2\theta_+\left({\mathcal F}_{0+}+{\mathcal F}_{0-}\right),\\
 [{\mathcal F}_{TR}]&= \frac{M}{R-M}\cos\theta_+\left({\mathcal F}_{0+}+{\mathcal F}_{0-}\right),\\
 [{\mathcal F}_{R\varPhi}]&=-{\mathrm i}  \frac{R}{R-M}\left(R- M\sin\theta_+^2\right)\left({\mathcal F}_{0+}-{\mathcal F}_{0-}\right),\\
[{\mathcal F}_{R \bot}]&=[{\mathcal F}_{T \varPhi}]=0.
\end{split}
\end{align}
Using equations \eqref{eq:discontinuity_electrical_monopole} and \eqref{eq:discontinuity_magnetic_monopole}--\eqref{eq:discontinuity_dipole} we observe again that for electric/magnetic charges the radial current is vanishing and that the electric and magnetic charges do rotate around the axis. The current is vanishing for $R\to\infty$. The total electric or magnetic charge of such a system will be infinite.
This will be different for the case of the field discussed in the next subsection.

We will now treat the case of electric monopoles and magnetic dipoles independently of the case of magnetic monopoles and electric dipoles. Afterwards the results can be superposed.

\paragraph{Electric monopoles or magnetic dipoles}

This case is obtained for $E_{0+}=-E_{0-}=E_0$ and $H_{0+}=H_{0-}=H_0$, together with $A^{(e)}_{t0+}=A^{(e)}_{t0-}$ and $A^{(e)}_{\varphi 0+}=A^{(e)}_{\varphi 0-}$. This leads to a surface current
\begin{align}\label{eq:current_electric_monopoles_homogeneous}
\begin{split}
 s_{(e\mathpzc{Mo})}^{T}&=-\frac{E_{0}}{2\pi}N \left(-1+\frac{M}{R}\sin^2\theta_+\right),\\
 s_{(e\mathpzc{Mo})}^{R}&= 0,\\
 s_{(e\mathpzc{Mo})}^{\varPhi}&=-\frac{H_0}{2\pi} N\frac{M}{R^2}    \cos\theta_+ . 
\end{split}
\end{align}
In the classical case $M=0$ the charges are at rest with a charge density equal to the first factor in the first equation. The discontinuities in the magnetic potential and the tangential components of the dual of the Maxwell tensor are in this case understood as being caused by the discontinuities of the orthogonal components of the Maxwell tensor and the presence of the electric monopole layer and, hence, the impossibility to introduce a magnetic potential globally. Looking at the classical case $M=0$, the principal problem mentioned after equation \eqref{eq:line_integral} becomes apparent when dealing with fields which are not falling off sufficiently fast at infinity. The axial current vanishes in this limit and thus cannot cause the magnetic field. The existing magnetic field can be explained by ``magnetic charges or electric currents at infinity''. Therefore, the disk is not the only source of the external field. This problem does not occur for fields which are falling off sufficiently fast. Such are discussed in the next example. However, for completeness we give here the 4-current provided that the discontinuities are interpreted as the result of a magnetic dipole layer according to equation \eqref{eq:discontinuity_dipole}:
\begin{align}\label{eq:current_magnetic_dipole_homogeneous}
\begin{split}
 s_{(m\mathpzc{Di})}^{T}&=\frac{H_0}{2\pi}\frac{M R}{R-2M}\cos\theta_+,\\
 s_{(m\mathpzc{Di})}^{R}&=0,\\
 s_{(m\mathpzc{Di})}^{\varPhi}&=\frac{E_{0}}{4\pi}.
\end{split}
\end{align}
Here the constants $A^{(m)}_{T\pm}$ and $A^{(m)}_{\varPhi0\pm}$ are chosen such that the current is not singular at the axis and the dipole density vanishes at infinity. 

Analogously, we can study disks endowed with a magnetic charge density or electric dipole density by setting $E_{0+}=E_{0-}=E_0$ and $H_{0+}=-H_{0-}=-H_{0}$. The results are very similar to \eqref{eq:current_electric_monopoles_homogeneous} and \eqref{eq:current_magnetic_dipole_homogeneous}; they can be obtained by a substitution $E_0\to H_0$ and $E_0\to-H_0$ into \eqref{eq:current_electric_monopoles_homogeneous} and \eqref{eq:current_magnetic_dipole_homogeneous}.

\subsection{Disks generated by point charges}

The question whether a field is generated solely by disks or also by sources at infinity is circumvented if a solution is chosen such that it falls off sufficiently fast at infinity. We now consider the electromagnetic field produced by a point charge $e$ situated in an arbitrary position $(r_0,\theta_0,\varphi_0)$. The electric 4-potential for such a point charge was given in \cite{Bicak_1977}, and in closed form by Linet in \cite{Linet_1976}. It reads\footnote{The different sign in the potential has its origin in the exchange of the indices of the Maxwell tensor, cf. footnote \ref{footnote:convention}.}:
\begin{align}\label{eq:fourpotentialpointcharge}
\begin{split}
A^{(e)}_t&=-\frac{M e}{r r_0}-\frac{e}{D r r_0}\left(\left(r-M\right)\left(r_0-M\right)-M^2\lambda\right),\\
A^{(e)}_r&=A^{(e)}_\theta=A^{(e)}_\varphi=0,\\
\lambda&=\cos\theta\cos\theta_0+\sin\theta\sin\theta_0\cos(\varphi -\varphi_0),\\
D&=\left((r-M)^2 + (r_0-M^2)-M^2\right.\\
 &\left. -2(r-M)(r_0-M)\lambda+M^2\lambda^2\right)^{\frac 1 2}.
\end{split} 
\end{align}
We consider two different test fields in the Schwarzschild spacetimes: the field produced by a point charge at $(r_+,\theta_+,\varphi_+)$ and the field produced by a point charge at $(r_-,\theta_-,\varphi_-)$.
In the spacetime with the first test field we make a cut at such $z=z_0$ that the black hole and the point charge are {\em{below}} the cut. For the second test field the cut is made at $z=-z_0$ such that the charge and the black hole are {\em{above}} the cut. After identifying the two hypersurfaces $z=\pm z_0$ there is no black hole or point charge in the spacetime, rather a massive disk with electromagnetic sources. However, the electromagnetic field outside the disk and thus the sources can be understood using the field lines in the ``original'' spacetime for the ``original'' test field, i.e., the Schwarzschild black hole spacetime with a point charge. This point of view is employed several times in the following; e.g., the charge density of the disk is explained by referring to the ``original'' black hole and its polarization.

The fields of the two point charges can be obtained from the 4-potential \eqref{eq:fourpotentialpointcharge} in a straightforward way and so also the jumps. In order to obtain a layer endowed with either only charges or dipoles we have to require that the point charges have to be located symmetrically in the original spacetime, i.e., $r_{0+}=r_{0-}=r_0,~\varphi_{0+}=\varphi_{0-},~\theta_{0+}=\pi-\theta_{0-}$, as well as that the charges are either equal, $e_+=e_-$, or opposite, $e_+=-e_-$. Because of the axially symmetry of the  spacetime we can set $\varphi_{0+}=0$. The jumps evaluate to
\begin{align}\label{eq:jumpspointcharge}
\begin{split}
[A^{(e)}_T]&=\frac{(e_2-e_1)}{r_{0} R D}  \left((R-M)(r_{0}-M)+M D-M^2 \lambda \right),\\
[A^{(e)}_R]&=[A^{(e)}_\varPhi]=[A^{(e)}_\bot]=[F_{R\bot}]=[F_{\varPhi\bot}]=0,\\
[F_{T\bot}]&=\frac{(e_1+e_2) (\xi_2-2 M)}{r_1 \sqrt{(M-\xi_2)^4-z_0^2 M^2} \xi_2^2 D^3} \times\\
 &\left[z_0 \left((M-r_1)\left((M-\xi_2)^2 \xi_2-M D^2\right)-MD^3\right)\right.\\
&\left.+z_0 \left((\xi_2-M)\xi_2(M^2+(M-r_1)^2)+M^2 D^2\right) \lambda\right.\\
&\left.+z_0 M^2 (M-r_1) \xi_2 \lambda ^2\right.\\
&\left.+(2 M-r_1) r_1 \sin \theta_+ (M-\xi_2)^2 \xi_2 \lambda_{,\theta}\right],\\
\end{split}
\end{align}
Note that functions $D$ and $\lambda$ have to be evaluated at the respective $\Sigma_\pm$ with the respective point charge. However, it holds that $\lambda(\theta_{+},\theta_{0+})=\lambda(\pi-\theta_{+},\pi-\theta_{0+})$, so the same holds for $D$. Therefore, functions $D$, $\lambda$ and $\lambda_{,\theta}$ should be read as functions with the argument $r=R,~\theta=\theta_+=\arccos\frac{z_0}{R-M},~\varphi=\varPhi,~r_0=r_{0+},~\theta_{0}=\theta_{0+},~\varphi_{0+}=0$. The jumps of the tangential components of the Maxwell tensor can be inferred from the jumps of the 4-potential. We can now discuss two cases -- a monopole layer and a dipole layer.

\paragraph{Electric monopoles or magnetic dipoles}

\begin{figure}[h]
\includegraphics[scale=0.205]{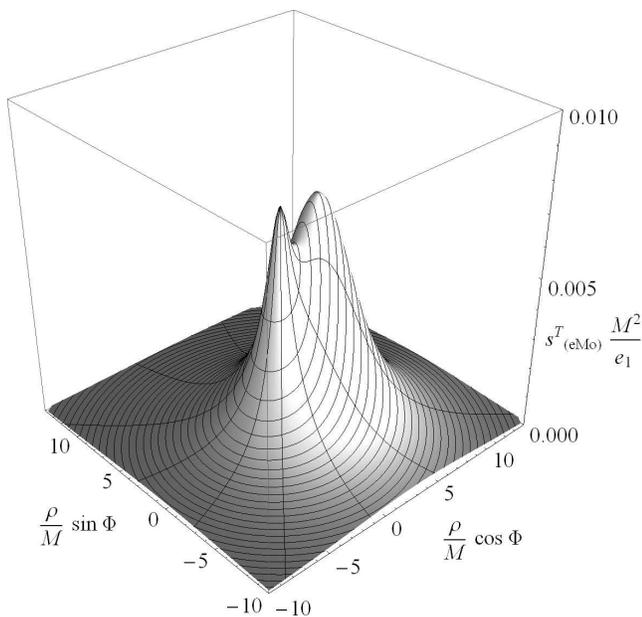}
\caption{\label{fig:chargedensity}The time component of the surface current $s_{(e\mathpzc{Mo})}^a$ (i.e., the charge density) endowed with electric charges for the parameters $r_{0+}=5.1 M$, $\theta_{0+}=0.7\pi$, $z_0=1.7 M$.}
\end{figure}

In order to obtain continuous tangential components of the 4-potential we have to set $e_1=e_2$. Then the surface 3-current can be read off \eqref{eq:jumpspointcharge} and \eqref{eq:discontinuity_electrical_monopole}. The only non-vanishing component is $s_T$. However, it is possible to consider two counterrotating streams with an equal charge, cf. with the underlying matter currents in the Schwarzschild disk \cite{Bicak_1993}. This would of course change the charge density seen by a comoving observer. There are several parameters governing the behavior of the solution: the cut parameter $z_0$, the charge $e_1$ which acts as scaling, and the position of the two charges $\{r_0,\theta_{0\pm},0\}$. In general, there is one maximum associated with the position of the charge $e_1$ as in classical electrodynamics, and there is also the second maximum due to the influence of the black hole, as depicted in Fig. \ref{fig:chargedensity}. Although for $\theta_{0+}=0$ an axially symmetric distribution is obtained, so, only one maximum is present in this case.
In the general case the first maximum lies at $\varPhi=0$ and the second at $\varPhi=\pi$, i.e., on opposite the side of the black hole in the ``original'' spacetime. 
The second maximum can be understood using the membrane paradigm \cite{Thorne_1986} (alternatively by discussing the boundary conditions at the horizon \cite{Punsly_2008}). Interpreting the horizon as a conducting sphere, a polarization is to be expected due to the field of the test charge. This will lead to a fictitious charge density at the horizon, cf. \cite{Thorne_1982}, as follows:
\begin{align}\label{eq:fictuouschargedensity}
\begin{split}
 \sigma_{H\pm}&= e_1 \frac{M(1+\lambda_\pm^2)-2(r_0-M)\lambda_\pm}{8\pi r_0 (r_0-M(1+\lambda_\pm))^2},\\
\lambda_\pm&=\pm\cos\theta\cos\theta_{0+}+\sin\theta\sin\theta_{0+}\cos\varphi,
\end{split}
\end{align}
where the upper sign denotes the induced charge density for the charge $e_1$ at $\{r_0,\theta_{0+},0\}$ and the lower for the charge $e_1$ at $\{r_0,\pi-\theta_{0+},0\}$. 
In the following we discuss only the $+$ case, the other one follows from the reflection symmetry. Assuming $e_1>0$, the area of the conducting sphere characterized by
\begin{align}
 r_0-\left(r_0^2-2 M r_0\right)^{\frac 1 2}\leq M (1+\lambda_+)\leq  r_0+\left(r_0^2-2 M r_0\right)^{\frac 1 2}
\end{align}
is negatively charged. The opening angle $\alpha_{crit}$ as seen from the test charge $e_1$ for this area was described in \cite{Hanni_1973}. There it was also discussed, that the field lines emanating from $\{r_0,\theta_{0\pm},0\}$ with an angle $\alpha\leq\alpha_{crit}$ are bent towards the horizon and cross it eventually. Field lines starting at $\alpha>\alpha_{crit}$ are first bent towards the horizon due to the opposite sign of its charge density and then bent away because of the change of sign in the polarization density. This leads to an increase/decrease of the tangential/normal components of the electric field in the disk close to the axis of the black hole facing $e_1$.  On the other side of the black hole the normal/tangential components of the electric field in the disk are increased/decreased. Thus in general, two maxima for the charge density are obtained on opposite sides of the axis. For the dipole density also two maxima are to be expected but both are lying on the side of the black hole facing $e_1$.

The surface charge current in $\Sigma$ behaves for $R\to\infty$ like
\begin{align}\label{eq:asymptoticchargedensity}
s_{(e\mathpzc{Mo})}^T(R,\varPhi)\sim \frac{e_1 (z_0+(2 M-r_{0+}) \cos\theta_{0+})}{2\pi R^3}.
\end{align}
The fall off is sufficiently fast to permit the definition of the total charge which can of course be read off from the unchanged asymptotic behavior of the field and thus is still $e_1$. Having fixed $r_{0+}$, the parameter $z_0$ can be used to slow down the decrease of the charge density as can be seen from \eqref{eq:asymptoticchargedensity}, but since the total charge must remain the same, the charge gets only ``smeared out''. 

\paragraph{Dipole disk}

\begin{figure}[htb]
\includegraphics[scale=0.17]{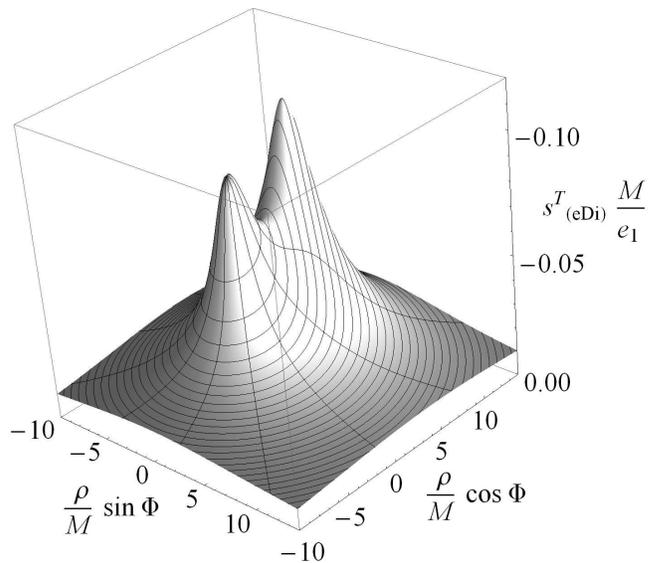}
\caption{\label{fig:dipoledensitypointcharge}The time component of the surface current $s_{(e\mathpzc{Di})}^a$ of the disk endowed with electric dipoles for the parameters $r_{0+}=6.5 M$, $\theta_{0+}=0.48\pi$, $z_0=1.7 M$.}
\end{figure}

To obtain continuous normal components of the Maxwell tensor one has to choose $e_1=-e_2$; the surface current is given by \eqref{eq:jump_in_potential_DP} and \eqref{eq:jumpspointcharge}. Again, the surface current allows two interpretations: the distribution is static or it consists of two counterrotating streams. The same parameters arise here as in the last case and the generic behavior for some specific values is depicted in Fig. \ref{fig:dipoledensitypointcharge}. The two maxima can again be understood on the grounds of the membrane paradigm as described above. 
The asymptotic behavior of the dipole density is
\begin{align}
s_{(e\mathpzc{Di})}^T(R,\varPhi)\sim -\frac{e_1 }{2\pi R}.
\end{align}

The relation between a monopole distribution and a dipole distribution is illustrated in the following\footnote{A more common fact is that magnetic dipoles can be interpreted as generated by a current of electric charges. For brevity we consider here the analogous case of electric dipoles and magnetic charges.}. Let us consider the electric 4-potential and the jumps in the tangential components of the Maxwell tensor as produced from the jumps in the normal components of the dual of the Maxwell tensor, i.e., of a magnetic charge density. If we remove the $\delta-$distribution terms of the field, we obtain a field which is generated by a magnetic current which satisfies
\begin{align}
\begin{split}
 s_{(m\mathpzc{Mo})}^T&=0,\quad s_{(m\mathpzc{Mo})}^R=-\epsilon^{(3)TR\varPhi}s_{(e\mathpzc{Di})T,\varPhi},\\
 s_{(m\mathpzc{Mo})}^\varPhi&=\epsilon^{(3)TR\varPhi}s_{(e\mathpzc{Di})T,R}.
\end{split}
\end{align}
As stated in section \ref{subsec:DiscontinuitiesGeneral} for the general case, it is obvious here that the continuity equation is also satisfied for the magnetic surface current. The magnetic charge density of this current is vanishing which can be interpreted as two currents with opposite charges, one of them at rest for example. Since the field falls off sufficiently fast and no total charge is present this is the sole source of the field.  

It is again clear from the symmetry of the Maxwell equations that the calculations of this section can be repeated for a magnetic point charge in order to obtain a magnetic charge density or a magnetic dipole density.

From our analysis it follows that similarly we could endow disks with test charges and dipoles which produce Kerr spacetimes \cite{Bicak_1993a}.
\begin{acknowledgments}
We thank Tom\'a\v s Ledvinka for helpful discussions. JB acknowledges the partial support from Grant No. GA\v CR 202/09/0772 of the Czech Republic, of Grants No. LC06014
and No. MSM0021620860 of the Ministry of Education. NG was financially supported by the Grants  No. GAUK. 22708 and No. GA\v CR 205/09/H033. JB and NG are also grateful to the Albert Einstein Institute in Golm for the kind hospitality.
AC\,G-P acknowledges the hospitality of the Institute of Theoretical Physics, Charles University (Prague) and the financial support from COLCIENCIAS, Colombia. 
\end{acknowledgments}

\end{document}